\documentclass[10pt,compsoc,cspaper,fleqn]{IEEEtran}

\usepackage{cite}

\ifCLASSINFOpdf
\else
\fi
\usepackage{amsfonts,amssymb,amsmath}
\usepackage{float}
\usepackage{subfigure}
\usepackage{color}
\usepackage{graphicx}
\usepackage{multirow}
\usepackage{alltt}
\definecolor{tabgrey}{rgb}{0.8,0.8,0.8}
\usepackage[mathscr]{eucal}
\usepackage{arydshln}
\usepackage{stmaryrd}
\usepackage{algorithmicx}
\usepackage{algpseudocode}

\usepackage[table,usenames,dvipsnames]{xcolor}
\usepackage[T1]{fontenc}
\usepackage[tableposition=top]{caption}
\usepackage{rotating}
\usepackage{color}
\usepackage{booktabs,array,dcolumn}
\usepackage{multirow}
\usepackage{varwidth}
\usepackage{xcolor,colortbl}
\usepackage{footmisc}

\definecolor{LightCyan}{rgb}{0.88,1,1}
\definecolor{LightGray}{gray}{0.9}
\newcolumntype{a}{>{\columncolor{LightCyan}}r}
\newcolumntype{b}{>{\columncolor[gray]{.9}[5pt]}r|}
\usepackage{hhline}

\newcommand{\nop}[1]{}

\newcommand{\hide}[1]{\hspace*{-5pt}~}
\floatstyle{ruled}
\newfloat{algorithm}{thp}{loa}
\floatname{algorithm}{Algorithm}

\usepackage{caption}
\usepackage{paralist}

\usepackage{footnote}
\makesavenoteenv{tabular}
\makesavenoteenv{table}

\usepackage{url}

\usepackage{balance}

\begin{document}
\title{Unreported links between trial registrations and published articles were identified using document similarity measures in a cross-sectional analysis of ClinicalTrials.gov}

\author{Adam G. Dunn\textsuperscript{1},
		Enrico Coiera\textsuperscript{1},
		Florence T. Bourgeois\textsuperscript{2,3}
\thanks{1. Centre for Health Informatics, Australian Institute of Health Innovation, Macquarie University, Sydney, Australia.}%
\thanks{2. Computational Health Informatics Program, Boston Children's Hospital, Boston, United States.}%
\thanks{3. Department of Pediatrics, Harvard Medical School, Boston, United States.}%
\thanks{Contact: adam.dunn@mq.edu.au}}

\IEEEtitleabstractindextext{%
\begin{abstract}
\newline
\textbf{Objectives:} Trial registries can be used to measure reporting biases and support systematic reviews but 45\% of registrations do not provide a link to the article reporting on the trial. We evaluated the use of document similarity methods to identify unreported links between ClinicalTrials.gov and PubMed.

\textbf{Study Design and Setting:} We extracted terms and concepts from a dataset of 72,469 ClinicalTrials.gov registrations and 276,307 PubMed articles, and tested methods for ranking articles across 16,005 reported links and 90 manually-identified unreported links. Performance was measured by the median rank of matching articles, and the proportion of unreported links that could be found by screening ranked candidate articles in order.

\textbf{Results:} The best performing concept-based representation produced a median rank of 3 (IQR 1-21) for reported links and 3 (IQR 1-19) for the manually-identified unreported links, and term-based representations produced a median rank of 2 (1-20) for reported links and 2 (IQR 1-12) in unreported links. The matching article was ranked first for 40\% of registrations, and screening 50 candidate articles per registration identified 86\% of the unreported links.

\textbf{Conclusions:} Leveraging the growth in the corpus of reported links between ClinicalTrials.gov and PubMed, we found that document similarity methods can assist in the identification of unreported links between trial registrations and corresponding articles.
\end{abstract}

}

\maketitle

\IEEEdisplaynontitleabstractindextext

%
\IEEEpeerreviewmaketitle

\IEEEraisesectionheading{\section{Background}\label{sec.intro}}

%
%
%
%

 

\IEEEPARstart{C}{linical} trial registries were established to track the conduct of clinical trials and make basic information about trials publicly available. A number of policies now mandate prospective registration for clinical studies of regulated interventions~\cite{Zarin2011,Zarin2005,McCray2000}. ClinicalTrials.gov is a US-based registry for clinical studies and is the largest single database of trial registrations. ClinicalTrials.gov also links registrations to published results by connecting to research articles indexed in bibliographic databases~\cite{Zarin2016,Zarin2008}. This linkage is achieved using a unique identifier (the NCT Number) for each study. Publishers may include the NCT number in the abstract or full text of published articles, and the metadata stored by PubMed, which provides access to a bibliographic database with information for more than 26 million biomedical articles.\\

While the introduction of trial registries has been invaluable for monitoring trial reporting, a substantial proportion of trials reports remain disconnected from their registrations. In a 2012 study examining the quality of linking in ClinicalTrials.gov, 44\% of registrations without linked publications were found to have corresponding published articles found by manual searches~\cite{Huser2012}. In a systematic review of studies that examine reported and unreported links between registrations and articles, the median proportion of registrations with reported links was 23\% and the median proportion of unreported links (those that required manual searches) was 17\%, with the remainder unpublished~\cite{Bashir2017}.\\

The quality of this linkage between bibliographic databases and trial registries affects the time it takes to measure reporting biases. This includes determining which clinical studies remain unpublished~\cite{Riveros2013,Baudart2016,Dickersin1987,Schmucker2014,Chalmers2013}, or comparing registered outcomes with what is reported in published articles~\cite{Bourgeois2010,Chan2004,Chan2005,Rising2008,Chan2008,Rasmussen2009,Rosenthal2013,Pranic2015,Scott2015}. Without comprehensive linking, research to evaluate reporting biases must instead rely on time-consuming manual searches to identify unreported links.\\

The presence of unreported links also limits the value of trial registries for systematic reviews. If links between registrations and articles were comprehensive, registrations could be more effectively used to automate the identification of trials for inclusion in systematic reviews~\cite{OMaraEves2015,Tsafnat2014}, as well as provide early signals that a systematic review should be updated~\cite{Takwoingi2013,Shekelle2014,Pattanittum2012,Moher2008, Garner2016,Cohen2009,Cohen2012,Chung2012,Barrowman2003,Ahmadzai2013}.\\

Our aim in this study was to evaluate whether we could use information contained in the recorded links between ClinicalTrials.gov registrations and PubMed articles to help identify unreported links. The longer-term goal is to develop robust methods to identify all published research associated with a trial registration, whether or not links are provided in the registration record.

\section{Methods}\label{sec.methods}

In the following experiments, we use similarities in the text from registrations in ClinicalTrials.gov and articles in PubMed. Using the set of reported links as a baseline, we test a series of different methods to represent the text as features, and assign weights to each of the features. The resulting set of features are used to produce an automatic, weighted search query for use in PubMed, where the objective is to rank the matching article as high as possible in a list of candidate articles. The approach is expected to replace the need for an expert to construct search queries in PubMed for every registration without a link to a published article.\\

\subsection{Study data}\label{subsec.studydata}

We included trial registrations in ClinicalTrials.gov for trials that were received by ClinicalTrials.gov on or after 1 October 2007, were marked as completed, and described an interventional study. The date was selected to correspond to the passing of the Food and Drug Administration Amendments Act of 2007, which expanded registration requirements for studies registered with ClinicalTrials.gov. A final search of ClinicalTrials.gov was conducted on April 14 2017. Data extracted from registrations included titles, trial summaries, and conditions studied.\\

We next selected all articles indexed in PubMed that reported a clinical trial and were published on or after October 1 2007. Articles were assumed to be reporting the results of trials if they included a ClinicalTrials.gov NCT Number as a secondary source identifier or listed ``clinical trial'', ``controlled clinical trial'', or ``randomized controlled trial'' as a publication type, and did not include ``meta-analysis'' or ``review'' as a publication type. A final search was conducted on April 14 2017. Data extracted from each PubMed article entry included the title text, abstract text, and any NCT Numbers stored as a secondary source identifier in the metadata. Where PubMed entries included NCT Numbers as secondary source identifiers, we described these as \textit{reported links} and created a dataset comprising the set of registrations with known links from one or more articles.\\

We then created a second set of registrations for testing, comprising 90 registrations that had \textit{unreported links} to trial articles identified by manually checking 200 registrations with trial completion dates between January 1 2007 and December 31 2015. The 200 registrations had no reported links to trial articles in PubMed at the time of the search. We manually searched PubMed and other bibliographic databases to identify articles that reported the results of the trials, following a search strategy previously described and common to studies examining outcome reporting biases~\cite{Bashir2017}. The search uses study design information, investigator names, locations, and other identifying features to search PubMed for the matching article. To confirm a match, we compared the number of participants, the study design and the length of the study, and any information about when and where the trial was undertaken. Where there were multiple matches, we selected the article published closest to the completion date of the trial.\\

\subsection{Feature representations and distance measures}\label{subsec.documentrep}

Data elements were extracted for each registry entry including brief title, detailed title, brief summary, detailed summary, and condition. Articles were represented by the text of the titles and abstracts in PubMed.\\

A \textit{term-based representation} of registrations and trial reports was created using the set of extracted words after removing punctuation. Using existing methods for extracting concepts from free text~\cite{Wallace2010,Matwin2010,Cohen2008,Ji2015,Hersh2000,Koopman2012,Zhang2007}, we also created a concept-based representation where each document was represented by the set of clinical concepts. These concepts were produced by MetaMap~\cite{Aronson2010}, which identifies concepts from the Unified Medical Language System (UMLS). We used a local MetaMap server that recognised concepts from the 2014AA UMLS release, and used MetaMap with word sense disambiguation. This method uses a term's context in a sentence to address the ambiguity of words with multiple meanings. The concept that was ranked highest by MetaMap for each phrase in the text was recorded.\\

We did not use information related to investigators, funding, or authors to allow unbiased testing of these document similarity measures. While investigator names, numbers of participants, and funding information are all useful for determining whether a published article describes a trial from a registration, it would need to be implemented differently from the terms and concepts used here and could introduce biases in the estimation of publication rates for studies with different funding sources.\\

For both the term and concept-based representations, we tested two ways to assign values to each of the features. A binary vector representation captured the presence or absence of a given feature (term or concept) in a document. A \textit{term frequency-inverse document frequency} (\textit{tf-idf}) vector representation was also calculated~\cite{Salton1988,Sparck1972}, given by the product of the number of times the feature appears in the document (log-transformed) and the inverse document frequency, which is given by the inverse of the proportion of documents in which the feature appears (also log-transformed). The log transformations are typically applied to the tf-idf calculation to help to ensure that individual terms or concepts that are used frequently in very few registrations and articles do not dominate the document similarity calculations described below.\\

We evaluated the performance of three standard pairwise measures of distance between the vectors representing registrations and the vectors representing trial reports. The \textit{normalised Euclidean distance} is given by the straight-line distance between the two vectors representing a registry entry and an article, divided by the number of features present in the article. The cosine similarity is the cosine of the angle between the two vectors representing a registry entry and an article. The \textit{Jaccard distance} is the number of features present in both a registration vector and article trial report vector, divided by the number of features present in either vector.\\

For each registry entry, we produced a ranked list of PubMed articles, created by sorting the set of all articles by their distance from the registration. Where a registry entry had known links from more than one trial report in PubMed, we determined the rank for the first linked trial report published after the completion date of the trial.

\subsection{Performance measures}\label{subsec.performancemeasures}

Performance was assessed using the final ranks of the linked (reported or unreported) trial reports, reported as median and interquartile range. We additionally determined the proportion of registrations for which the linked (reported or unreported) trial report was ranked first among all candidates, and the proportion of registrations for which the linked (reported or unreported) trial report was within the first 50 candidates (recall@50). We also reported the extension of this measure for recall after checking any number of candidates (recall@N), as a visual demonstration of the amount of manual effort required to identify a given percentage of previously unreported links in a cohort of registrations. The recall@N is the cumulative proportion of registrations for which the linked (reported or unreported) trial report can be found after having read the top N candidates, and is defined for values between 0 (where the proportion of found trial reports is always 0) and the total number of candidates (where the proportion is always 1).\\

Code for the document similarity methods are available online via a GitHub repository (https://github.com/pmartin23/tfidf). The repository includes scripts for using the document similarity to rank PubMed articles given a ClinicalTrials.gov NCT Number.\\

\section{Results}\label{sec.results}

\begin{figure*}[h]
\centering
\includegraphics[width=6in]{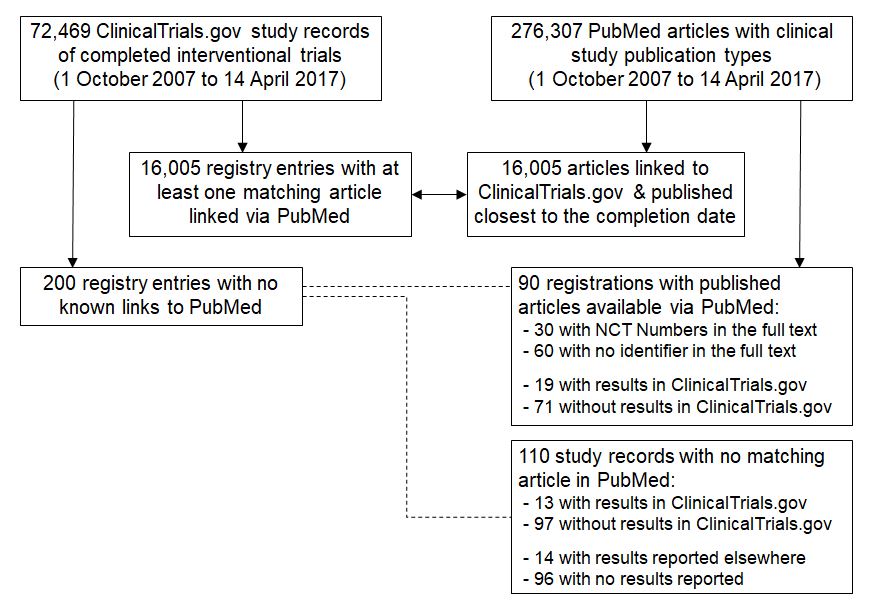}
\caption{The study data included 72,469 registrations from ClinicalTrials.gov and the metadata from 276,307 articles available in PubMed. A testing set of 200 registrations were manually reviewed and 90 of those were found to have unreported links to trial reports.}
\label{fig1}
\end{figure*}

We identified 72,469 registrations in ClinicalTrials.gov and 276,307 candidate trial reports in PubMed that met the inclusion criteria (Figure~\ref{fig1}). Among the registrations, 16,005 (22.1\%) had one or more links from trial reports published in the same period (Table~\ref{tab1}).

\begin{table}
\caption{Characteristics of 16,005 trials with completion dates after October 1 2007 and links from PubMed articles.}
\centering
\begin{tabular}{ |l|c| }
 \hline
 & \textbf{Count (\% of total)} \\
 \hline
 \textbf{Trial type} & \\
 ~Phase 0 \and 1 & 2,318 (14.5\%) \\
 ~Phase 2 & 2,931 (18.3\%) \\
 ~Phase 3 & 2,842 (17.8\%) \\
 ~Phase 4 & 2,091 (13.1\%) \\
 ~Unknown & 5,823 (36.4\%) \\
 \hline
 \textbf{Participants} & \\
 ~Under 50 & 4,946 (30.9\%) \\
 ~50-200 & 5,719 (35.7\%) \\
 ~Over 200 & 4,990 (31.2\%) \\
 ~Unknown & 98 (0.6\%) \\
 \hline
 \textbf{Funding} & \\
 ~Industry only & 4,759 (29.7\%) \\
 ~Mixed & 1,461 (9.1\%) \\
 ~No industry & 9,785 (61.1\%) \\
 \hline
 \textbf{Total} & 16,005 (100\%) \\
 \hline
\end{tabular}
\label{tab1}
\end{table}

In the set of 200 registrations, 90 were found to have unreported links to trial reports in PubMed. Of these, 33\% (30 of 90) included the NCT Number in the full text of the article but the identifier was not reported in the PubMed entry metadata or provided by investigators in ClinicalTrials.gov (Table~\ref{tab2}).

\begin{table}
\caption{Characteristics of 200 trials with completion dates after January 1 2007 and no reported links to PubMed.}
\centering
\begin{tabular}{ |l|c|c|c| }
 \hline
 & \multicolumn{1}{|p{1.6cm}|}{\centering \textbf{Count (\% of total)}}
 & \multicolumn{1}{|p{1.6cm}|}{\centering\textbf{Matching article in PubMed (\% of group)}}
 & \multicolumn{1}{|p{1.6cm}|}{\centering\textbf{NCT Number included in article (\% of group)}} \\
 \hline
 \textbf{Trial type} & & & \\
 ~Phase 0 \& 1 & 34 (17\%) & 12/34 (35\%) & 5/34 (15\%) \\
 ~Phase 2 &	52 (26\%) & 29/52 (56\%) & 7/52 (13\%) \\
 ~Phase 3 & 35 (18\%) & 15/35 (43\%) & 6/35 (17\%) \\
 ~Phase 4 & 30 (15\%) & 16/30 (53\%) & 6/30 (20\%) \\
 ~Unknown & 49 (24\%) & 18/49 (37\%) & 6/49 (12\%) \\
 \hline
 \textbf{Participants} & & & \\
 ~Under 50 & 88 (44\%) & 40/88 (45\%) & 13/88 (15\%) \\
 ~50-200 & 85 (42\%) & 36/85 (42\%) & 13/85 (15\%) \\
 ~Over 200 & 25 (12\%) & 14/25 (56\%) & 4/25 (16\%) \\
 ~Unknown & 2 (1\%) & 0/2 (0\%) & NA \\
 \textbf{Funding} & & & \\
 ~Industry only & 91 (46\%) & 33/91 (36\%) & 10/91 (11\%) \\
 ~Mixed & 17 (8\%) & 13/17 (76\%) & 5/17 (29\%) \\
 ~No industry & 92 (46\%) & 44/92 (48\%) & 15/92 (16\%) \\
 \hline
 \textbf{Total} & 200 (100\%) & 90/200 (45\%) & 30/200 (15\%) \\
 \hline
\end{tabular}
\label{tab2}
\end{table}

\begin{figure*}[h]
\centering
\includegraphics[width=7in]{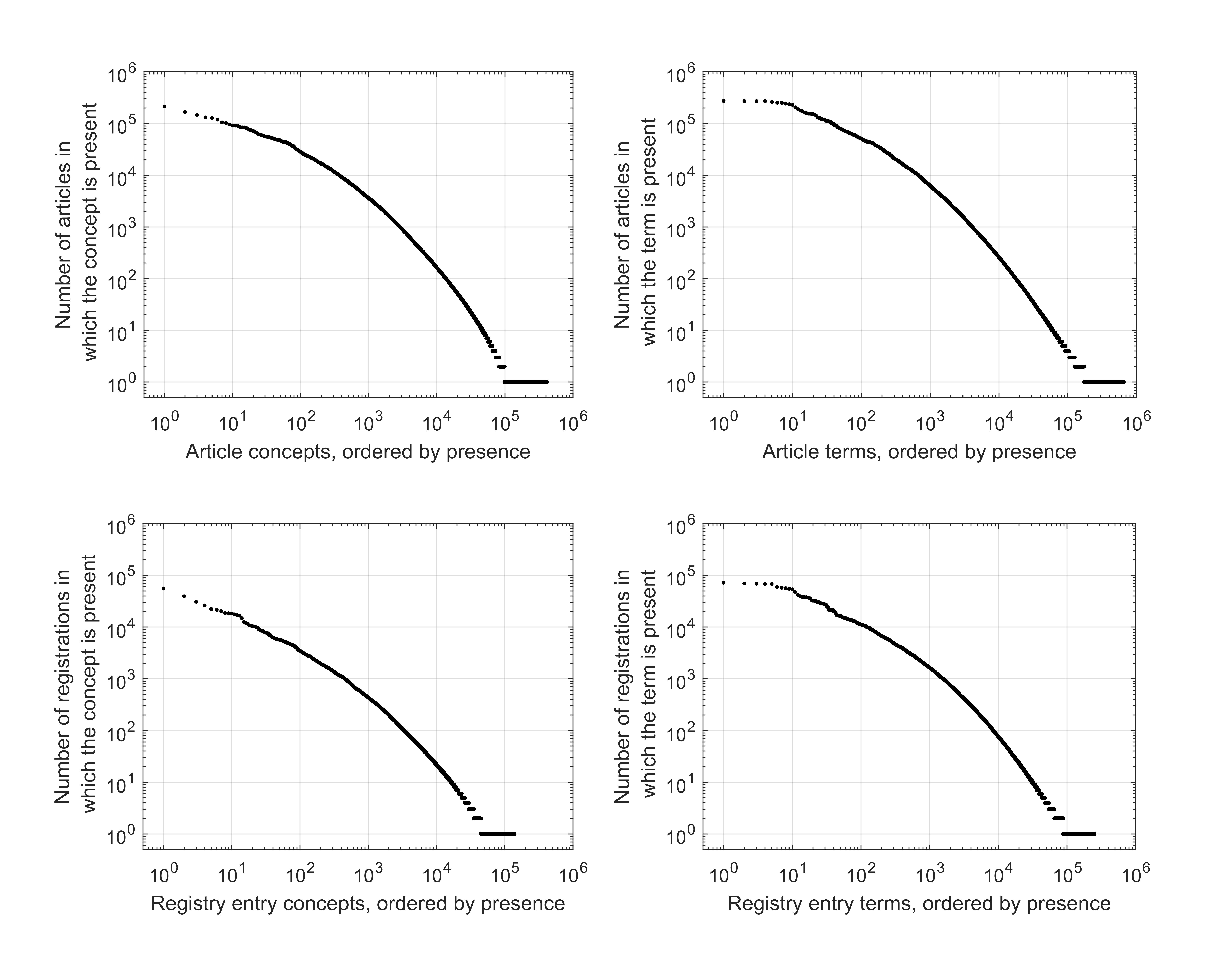}
\caption{The distribution of 490,814 concepts (left) and 789,848 terms (right) that were present in 276,307 articles (top) and the 72,469 registrations (bottom).}
\label{fig2}
\end{figure*}

The number of unique concepts found across both corpora was 490,814, and of these 46,777 were found more than once in both the registrations and the articles (Figure~\ref{fig2}). There were 789,848 unique terms across both corpora, and of these 78,856 were found more than once in both the registrations and the articles. After reducing the sets of concepts and features to include only those that were found more than once in both the registrations and articles, the number of remaining concepts varied between 2 and 273 (median 32) per registration, and between 2 and 296 (median 79) in articles. After applying the same process, the number of remaining terms per registration varied between 2 and 1,576 (median 108), and between 2 and 472 (median 142) for articles.\\

The median ranks produced by testing combinations of feature representations and distance measures on reported links varied substantially (Table 3). There were only small differences between the median ranks of the term-based and concept-based representations for the best performing combination, which in both cases used tf-idf weights and cosine distances. In both the term-based representation and the concept-based representation the tf-idf score and cosine distance measure placed the linked article first among all candidates for 40\% of registrations (ranked first from among 276,307 candidate trial reports), and within the top 50 ranked trial reports (ranked $\leq$50 from 276,307 candidate trial reports) for 83\% of the registrations.

\begin{table}
\caption{The effect of using terms or concepts on the performance by feature representation and distance measures in a set of 16,005 registrations with reported links to articles.}
\centering
\begin{tabular}{ |l|c|c|c| }
 \hline
   \multicolumn{1}{|p{2.0cm}|}{\centering \textbf{Feature representations}}
 & \multicolumn{1}{p{2.0cm}|}{\centering\textbf{Median rank (IQR)}}
 & \multicolumn{1}{p{1.3cm}|}{\centering\textbf{First-ranked candidate*}}
 & \multicolumn{1}{p{1.3cm}|}{\centering\textbf{Recall@50}} \\
 \hline
 \textbf{Terms (binary)} & & & \\
 ~Euclidean & 533 (8-7,900) & 17.2\% & 34.4\% \\
 ~Jaccard & 6 (1-160) & 34.8\% & 67.1\% \\
 ~Cosine & 4 (1-81) & 37.4\% & 71.9\% \\
 \hline
 \textbf{Terms (tf-idf)} & & & \\
 ~Euclidean & 363 (3-15,110) & 18.5\% & 40.2\% \\
 ~Cosine & 2 (1-20) & 40.2\% & 83.2\% \\
 \hline
 \textbf{Concepts (binary)} & & & \\
 ~Euclidean & 25 (1-1,679) & 28.8\% & 54.2\% \\
 ~Jaccard & 11 (1-191) & 28.3\% & 64.1\% \\
 ~Cosine & 6 (1-89) & 32.2\% & 70.9\% \\
 \hline
 \textbf{Concepts (tf-idf)} & & & \\
 ~Euclidean & 97 (2-7,230) & 23.1\% & 46.9\% \\
 ~Cosine & 3 (1-21) & 39.6\% & 82.3\% \\ 
 \hline
\end{tabular}
*The first ranked candidate was the article that reported the results of the trial in the registration.
\label{tab3}
\end{table}

\begin{figure*}[h]
\centering
\includegraphics[width=6.8in]{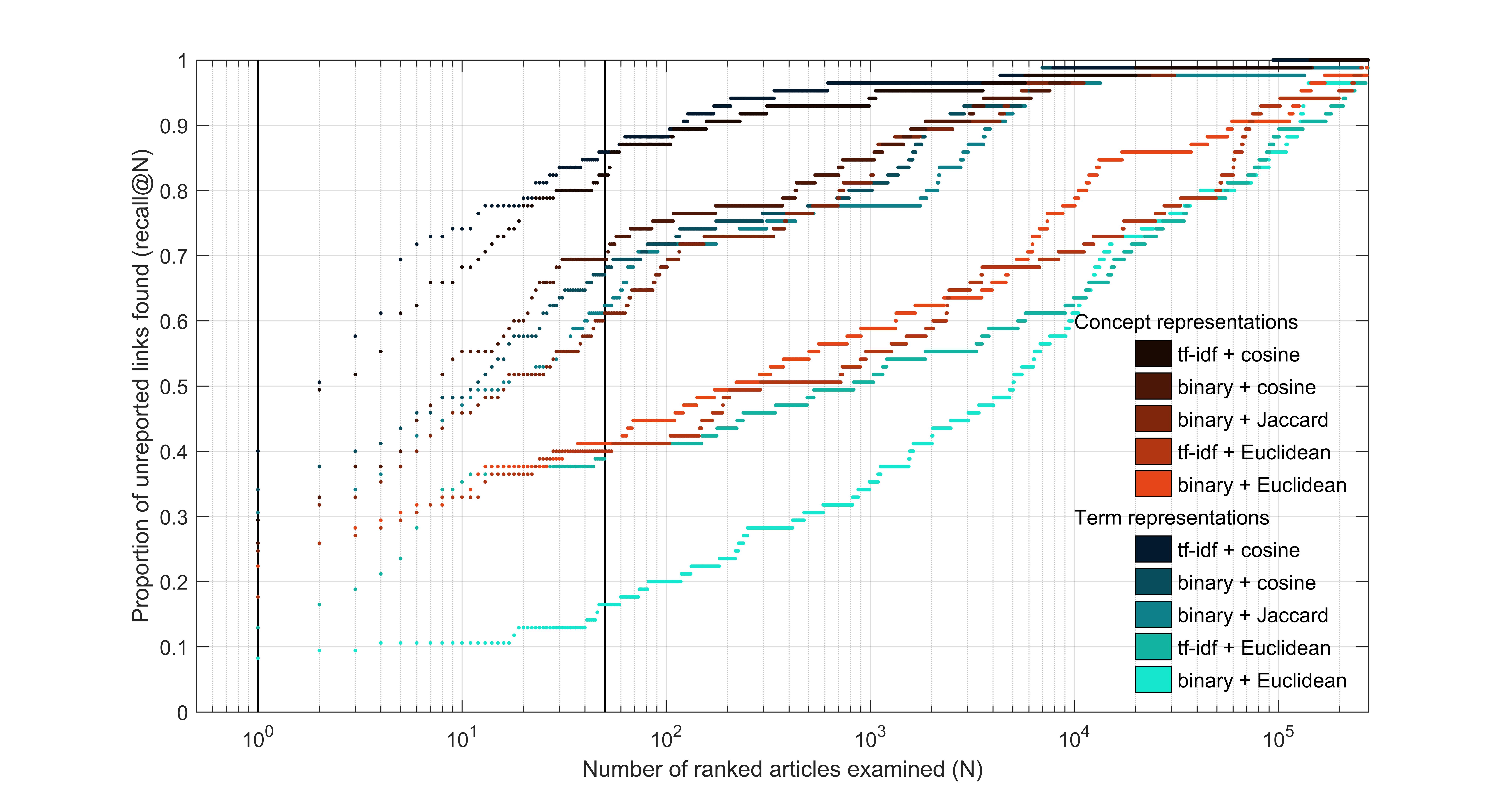}
\caption{Recall@N for term and concept based feature representations among the set of 90 registrations with unreported links to trial reports. The proportions of unreported links that were found by checking the first candidate (and first 50 candidates) are indicated by the two vertical lines.}
\label{fig3}
\end{figure*}

The results were similar in the 90 registrations with unreported links to trial articles, which were used to test the method in a practical setting. The best-performing combinations used the tf-idf score and the cosine distance measure (Table 4), and the term-based and concept-based representations produced similar results. The maximum recall@50 was 85.9\% for terms and 82.4\% for concepts. For other combinations of scores and distance measures, concept-based representations generally outperformed the equivalent term-based representations (Figure~\ref{fig3}).

\begin{table}
\caption{The effect of using terms or concepts on the performance by feature representation and distance measures in a set of 90 registrations with unreported links to trial reports.}
\centering
\begin{tabular}{ |l|c|c|c| }
 \hline
   \multicolumn{1}{|p{2.0cm}|}{\centering \textbf{Feature representations}}
 & \multicolumn{1}{p{2.0cm}|}{\centering\textbf{Median rank (IQR)}}
 & \multicolumn{1}{p{1.3cm}|}{\centering\textbf{First-ranked candidate*}}
 & \multicolumn{1}{p{1.3cm}|}{\centering\textbf{Recall@50}} \\
 \hline
 \textbf{Terms (binary)} & & & \\
 ~Euclidean & 5,030 (225-25,675) & 8.2\% & 16.5\% \\
 ~Jaccard & 16 (1-342) & 30.6\% & 62.4\% \\
 ~Cosine & 12 (1-206) & 34.1\% & 67.1\% \\
 \hline
 \textbf{Terms (tf-idf)} & & & \\
 ~Euclidean & 832 (6-29,048) & 12.9\% & 38.8\% \\
 ~Cosine & 2 (1-12) & 40.0\% & 85.9\% \\
 \hline
 \textbf{Concepts (binary)} & & & \\
 ~Euclidean & 221 (2-7,386) & 22.4\% & 41.2\% \\
 ~Jaccard & 17 (2-383) & 24.7\% & 61.2\% \\
 ~Cosine & 8 (1-92) & 25.9\% & 69.4\% \\
 \hline
 \textbf{Concepts (tf-idf)} & & & \\
 ~Euclidean & 291 (2-19,387) & 17.6\% & 40.0\% \\
 ~Cosine & 3 (1-19) & 29.4\% & 82.4\% \\ 
 \hline
\end{tabular}
\label{tab4}
\end{table}

\section{Discussion}\label{sec.discussion}

The results demonstrate that relatively simple representations across shared terms or concepts can be used to support the identification of unreported links between a trial's registration and the article reporting its results. For two in five registrations, the method ranks the matching article first among all candidates. For more than four in every five registrations, a user would only need to check 50 candidates to identify an unreported link to its matching article. 

\subsection{Comparisons with prior research}\label{subsec.comparisons} 

To the best of our knowledge no other methods have been proposed to automate the identification of unreported links between trials' registrations and the articles reporting their results. The most closely related research in the application domain is a method for identifying multiple publications from the same clinical trial~\cite{Shao2015}. Conceptually, the method is similar to concept-mapping that is used in certain search methods~\cite{Ide2007}, but differs because it produces what is effectively an automated and weighted query using the language of registrations rather than requiring users to author queries themselves. Related research in clinical epidemiology includes technologies that operate on bibliographic databases to support the screening of articles for inclusion in a systematic review~\cite{OMaraEves2015}. Recent developments included supervised, unsupervised, and hybrid methods~\cite{Ji2015,Ji2017,Miwa2014}, and several have taken advantage of semantic similarities between documents~\cite{Saha2016,Hashimoto2016}.\\

Our results on rates of reported and unreported links are consistent with previous observational studies examining ClinicalTrials.gov and PubMed. A series of studies showed that 27.8\% of 8,907 registrations for completed, interventional, Phase 2 or later trials were found to have one or more machine-readable links to PubMed, and 44\% of a sample of 50 registrations without known links were found to have matching trial reports~\cite{Huser2012,Huser2013a,Huser2013b}.

\subsection{Implications}\label{subsec.implications}

Our methods support novel approaches to the system-wide monitoring of trial reporting. A number of studies have used ClinicalTrials.gov to examine reporting biases, including both publication bias and outcome reporting bias~\cite{Dwan2013,Bashir2017}. In these studies, investigators relied on the manual identification of links between registrations and published articles to ensure that all published results were identified. This is a time-consuming and rate limiting step. By replacing the need for experts to construct search queries in PubMed for each registration without a link to a published article, our proposed approach could be used to reduce the expertise and time needed to identify unreported links between trial registrations and the articles that report their results. Because there are still a substantial proportion of links that remain unreported, and this proportion does not appear to have changed over time, the method is likely to be of continued value until existing registrations are comprehensively linked to their results.\\

In a systematic review of studies that examine both reported and unreported links between trial registrations and articles, the results indicated that there was a smaller proportion of unreported links when investigators started from a cohort of articles and tried to find links to registrations. This appeared to be because articles often included trial registry identifiers in the full text of articles that were not included in the metadata in bibliographic databases~\cite{Bashir2017}. We found a similar result: for 30 of 90 unreported links, the NCT Number was available in the full text of the article. All stakeholders, including the trial investigators, the journals, the bibliographic databases, and the organisations funding the trials could improve the number of links available in metadata, either by adding the information directly to ClinicalTrials.gov, or by ensuring the information is available in the metadata provided to PubMed.

The research also has implications for automating systematic review processes. Methods designed to help systematic reviewers identify articles for inclusion in systematic reviews often use machine learning to replicate human screening of articles. The best-performing methods in this area are able to reduce workload by between 30\% to 70\% with an estimated loss of 5\% of relevant studies~\cite{OMaraEves2015}. Comprehensive matching of registrations and trial reports could provide a more reliable and complete basis from which to develop methods for automating systematic review processes. Rather than relying on the minimal descriptions available for articles in bibliographic databases~\cite{Kiritchenko2010}, information from ClinicalTrials.gov could provide an earlier and more complete description of the trials and yield more accurate machine learning results for article screening and selection.

\subsection{Limitations and future research}\label{subsec.limitations}

This study has several limitations. In the set of unreported links, the manual search protocol may not have identified all published results. If the 90 articles we identified were easier to find by manual searches because they shared a larger number of terms, this may have over-estimated the performance in practice. However, the rate of publication for these entries was similar to prior reports~\cite{Bashir2017}. It is also possible that the set of reported links included articles that were not presenting the results of the trials (such as pilot studies, protocols, or secondary analyses), which could influence the performance relative to the manually-curated studies, which were all selected as the first articles presenting the results of the trials. To minimise the risk of including incorrect articles in the set of reported links, we included the reported link to the article that was published after the completion date of the trial and excluded studies that were published before the stated completion date.\\

The methods described here represent a series of baseline results that could be implemented directly in a process for finding unreported links, but could be improved in several ways. In particular, to ensure a fair comparison was made for the two approaches, we only included a limited selection of the fields available in ClinicalTrials.gov to represent the registrations. If we had included the names, affiliations, and countries of the investigators, it is likely that the performance of the document similarity method would have increased (but only for the term-based representation because these fields would not be recognised as UMLS concepts). Likewise, text from the full articles could have improved the performance of the method, but only when finding links to articles published in open access journals. Second, a further pre-processing step to reduce the number of candidate articles may improve the performance, especially where distance measures perform poorly. Finally, machine learning methods such as learning to rank approaches could be used to reduce, re-weight, or transform the sets of features by training them on the sets of reported links, and this may also yield improvements in performance given the volume and rate of growth in the number of reported links that can be used for training. Further research in this area should also consider how a user-friendly implementation of the methods tested here compare to a baseline in which experts and non-experts construct their own search strategies for PubMed.

\section{Conclusion}\label{sec.conclusion}

Information contained in clinical trial registries may be useful for monitoring trial reporting activities and synthesising evidence, but this type of surveillance remains limited due to the growing number of incomplete links between ClinicalTrials.gov and PubMed. Here we evaluate a method for partially automating the identification of published articles related to registrations. Our results demonstrate that document similarity methods can be used to replace the need to construct search strategies in a bibliographic database when trying to identify unreported links. While the approach does not replace the need to screen the resulting ranked set of candidates, it demonstrates the potential for automation in this space.

\ifCLASSOPTIONcompsoc
  \section*{Acknowledgments}
\else
  \section*{Acknowledgment}
\fi

This work was supported by the Agency for Healthcare Research and Quality (R03HS024798). The authors have no conflicts of interest to disclose.


\ifCLASSOPTIONcaptionsoff
  \newpage
\fi



%



\bibliographystyle{IEEEtran}
\bibliography{IEEEabrv,refbib}

%










\end{document}